\documentclass[%
 reprint,
 amsmath,amssymb,
 aps,
]{revtex4-2}

\usepackage{graphicx}%
\usepackage{dcolumn}%
\usepackage{bm}%
\usepackage{hyperref}%
\usepackage{amsmath,amssymb}
\usepackage[american]{babel}
\usepackage{hyperref} 
\usepackage{fontspec}
\usepackage{setspace}
\usepackage{xcolor}
\usepackage{amsmath}
\usepackage{subcaption}
\usepackage{float}
\usepackage{fontspec} 
\usepackage{verbatim}
\usepackage{bm}
\usepackage{url}
\usepackage{xr}
\usepackage{booktabs}
\usepackage{makecell}
\usepackage{xspace}
\hypersetup{colorlinks=true,linkcolor=blue,citecolor=blue, urlcolor=blue,pdfborder={0 0 0}   }
\usepackage{ragged2e}

\makeatletter
\long\def\@makecaption#1#2{%
  \par
  \vskip\abovecaptionskip
  \begingroup
  \small
  \justifying
  #1. #2\par
  \endgroup
  \vskip\belowcaptionskip
}
\makeatother

\begin{document}

\preprint{APS/123-QED}
\title{Fisher Information Metric as a model-free measure \\of proximity to criticality in neural systems}
\author{Yuewei Du$^{1}$}
\author{Alberto Liardi$^{2,3}$}
\author{Hardik Rajpal$^{3,5,6}$}
\author{Henrik Jeldtoft Jensen$^{3,4}$}  \email{h.jensen@imperial.ac.uk}

\affiliation{$^{1}$Department of Physics, Imperial College London, London, United Kingdom}
\affiliation{$^{2}$Department of Computing, Imperial College London, London, United Kingdom}
\affiliation{$^{3}$Centre for Complexity Science, Imperial College London, London, United Kingdom}
\affiliation{$^{4}$Department of Mathematics, Imperial College London, London, United Kingdom}
\affiliation{$^{5}$I-X Centre for AI in Science, Imperial College London, London, United Kingdom}
\affiliation{$^{6}$Department of Electrical and Electronic Engineering, Imperial College London, London, United Kingdom}
\date{\today}%

\begin{abstract}
Critical phenomena are widespread across many disciplines and have recently become a topic of deep interest in the study of biological and artificial neural networks.
A distinct signature of criticality is the emergence of avalanches with power-law-distributed sizes and durations.
However, empirically estimating the critical exponents remains challenging, and their interpretation is often model-dependent. In this work, we demonstrate how the Fisher Information Metric (FIM), a measure of generalized susceptibility, provides a comprehensive, model-agnostic characterization of the critical region in neural systems. 
We validate this approach across models of increasing biological complexity, from prototypical branching processes to spiking and whole-brain models, showing that FIM of each model's control parameter reliably tracks the system's degree of criticality. 
To emulate the study of real-world systems, where the control parameter is unknown, we additionally calculate FIM of the observed branching ratio of neural activity. The resulting FIM peaks where activity growth and decay balance, with the peak sharpening as the system approaches criticality. 
Hence, FIM peak width and height yield continuous, model-free readouts of proximity to criticality without requiring knowledge of the true control parameter, offering a robust tool for probing criticality in neural systems. 
\end{abstract}
\maketitle

\section{Introduction}\label{intro}
Criticality is a hallmark of complex systems across many domains, from earthquakes and forest fires to ecosystems and financial markets, marking a delicate balance between order and disorder at which systems display maximal sensitivity, long-range correlations, and rich, scale-free dynamics~\cite{jensen1998self, christensen2005complexity, sethna2021statistical, stanley1987introduction}. For several decades, evidence has accumulated suggesting that neuronal systems, such as the brain, operate close to such a critical regime~\cite{beggs2012being, RevModPhys.90.031001, beggs2012being, Wil19, hengen2025criticality, Meshulam_2025, chialvo2010emergent}. One of the most striking observations is the emergence of intermittent bursts of activity across scales, known as neuronal avalanches, whose sizes and durations exhibit scale-free distributions with consistent critical exponents in neuronal cultures~\cite{beggs2003neuronal, friedman2012universal, gireesh2008neuronal, shriki2013neuronal}, local field potentials (LFPs)~\cite{petermann2009spontaneous}, mesoscale calcium imaging~\cite{rajpal2026synergy}, and whole-brain EEG and fMRI recordings~\cite{tagliazucchi2012criticality,expert2010self}. Operating near this specific regime appears to be functionally advantageous: criticality marks the boundary between order and disorder at which long-range spatiotemporal correlations emerge~\cite{jensen2021critical, expert2010self}, dynamic range is maximized~\cite{kinouchi2006optimal}, and both information transmission~\cite{Ito_1994, Birdsey_2017, shew2011information} and computational capabilities~\cite{Boedecker_2011} are enhanced. Because the real brain lacks an obvious external control parameter, self-organized criticality (SOC) has been proposed to explain how systems may have naturally evolved towards a critical state~\cite{per1987self, bak1989earthquakes, bak1990forest, PhysRevLett.68.1244,jensen1998self,pruessner2012SOC}.
However, despite extensive experimental evidence and theoretical models, the \textit{criticality hypothesis}, i.e.\ whether neural systems operate at criticality, remains a debated topic~\cite{o2022critical, zimmern2020brain}. Part of this difficulty stems from the fact that we still lack precise, reliable methods for estimating the system's proximity to criticality, leaving open the question of how best to measure it in practice.

Various methods have been developed to address this challenge. A widely used approach identifies neuronal avalanches and compares their size and duration statistics against the mean-field branching process universality class~\cite{beggs2003neuronal, pruessner2012SOC, tagliazucchi2012criticality, haimovici2013brain, haldeman2005critical}. However, the inferred critical exponents depend sensitively on methodological choices such as activity thresholding and time binning~\cite{beggs2003neuronal, tagliazucchi2012criticality, petermann2009spontaneous}, and approximate power-law statistics can arise without true SOC~\cite{bedard2006does, touboul2010can, Touboul_2017, PRXLife.3.013013}. Other approaches focus on estimating the branching parameter directly from consecutive activity ratios~\cite{beggs2003neuronal, shriki2013neuronal, wilting2018inferring, zeraati2023intrinsic, poil2008avalanche}, but this presumes neural activity is well described by the uncorrelated branching process, an assumption at odds with the long-range correlations that are themselves a hallmark of criticality~\cite{jensen2021critical, expert2010self, rajpal2026synergy,lombardi2023beyond}. More recently, the temporal renormalization group approach has been applied directly to neural time series data, offering a principled, scale-free measure of proximity to criticality without relying on avalanche statistics~\cite{sooter2025defining, sooter2024cortex}. This elegant framework, however, requires fitting an explicit dynamical model (e.g.\ an autoregressive process) to the data, hence making it dependent on the chosen model class and its ability to capture the system's dynamics. 
To summarize, existing approaches of measuring proximity to criticality are constrained by the tension that model-specific methods offer clear interpretability but limited generality, while model-agnostic measures often sacrifice a direct link to the underlying dynamics. 

To address this gap, we consider the Fisher Information Metric (FIM), an information-geometric measure that quantifies the local sensitivity of a system's probability distribution to variations in its control parameters~\cite{Cover_Thomas_1991, amari2000methods, amari2016information}. This characteristic response makes FIM a useful estimator to quantify critical behavior~\cite{janke2004information}. In the statistical mechanics terminology, FIM is a measure of generalized susceptibility, measuring the response of the system to parameter variations~\cite{Crooks2012,hidalgo2014information}, diverging and peaking at the critical point in infinite and finite systems, respectively. For order parameters derivable from thermodynamic potentials, FIM elements are equal to the rate of change of the corresponding order parameter with respect to the relevant control parameter~\cite{prokopenko2011relating}. Moreover, for dynamics of finite networks, FIM has been shown to encode topology-dependent structural transitions, such as changes in the number of synchronized clusters near the edge of chaos~\cite{kalloniatis2018fisher}. Nevertheless, current non-parametric estimations of FIM require access to the control parameter of the data-generating model~\cite{PhysRevE.93.023301}. 

In this work, we develop a procedure that enables FIM to assess proximity to criticality in neural systems without knowledge of the true control parameter.
We apply this framework to three models of increasing biological complexity: the mean-field branching process~\cite{athreya2012branching, harris1963theory}, a model of spontaneously spiking neurons on a network~\cite{li2020tuning, larremore2014inhibition}, and a whole-brain model of BOLD signals on a human connectome~\cite{deco2013resting, deco2014local, herzog2024neural}. For each system, we adopt a twofold approach, computing FIM with respect to both the model's theoretical control parameter and the observed branching ratio, which measures the ratio of activity between consecutive time points. The second approach is motivated by the fact that in real systems the true control parameter is, in general, inaccessible or entirely unknown, whereas consecutive activity ratios can be directly measured from spike recordings and are widely used as empirical proxies for criticality~\cite{beggs2003neuronal, shriki2013neuronal, 
wilting2018inferring, zeraati2023intrinsic, poil2008avalanche}. Our results first clarify the dynamical significance of the observed branching ratio and its non-trivial relationship to proximity to criticality, and then demonstrate the validity of FIM as a practical, model-free estimator of criticality across all considered scenarios.

\begin{table*}[ht]
\centering
\setlength{\tabcolsep}{10pt} 
\renewcommand{\arraystretch}{1.2}
\begin{tabular}{ccc}
\Xhline{1.2pt}
\textbf{Model} & \textbf{Theoretical Control Parameter} & \textbf{Empirical Control Parameter} \\
\hline
Branching process & Average branching ratio $\mu$ & Generation ratio $\sigma=Z(t+1)/Z(t)$\\
Spiking neural network & Largest eigenvalue $\lambda$ & Branching ratio $\sigma=S(t+1)/S(t)$ \\
Whole brain model & Global coupling strength $g$ & Branching ratio $\sigma= \hat{S}(t+1)/\hat{S}(t)$ \\
\Xhline{1.2pt}
\end{tabular}
\caption{Summary of models and corresponding control parameters for calculating FIM (Eq.~\ref{Eq:fim_definition}). 
Here $Z(t)$ denotes the number of nodes in generation $t$ and $S(t) $ denotes the number of firing nodes a time $t$, finally, $\hat{S}(t)$ denotes the spatial average of the activity across brain regions (see text for details).}
\label{tab:control_param_summary}
\end{table*}

\begin{figure*}[ht]
    \includegraphics[width=\linewidth]{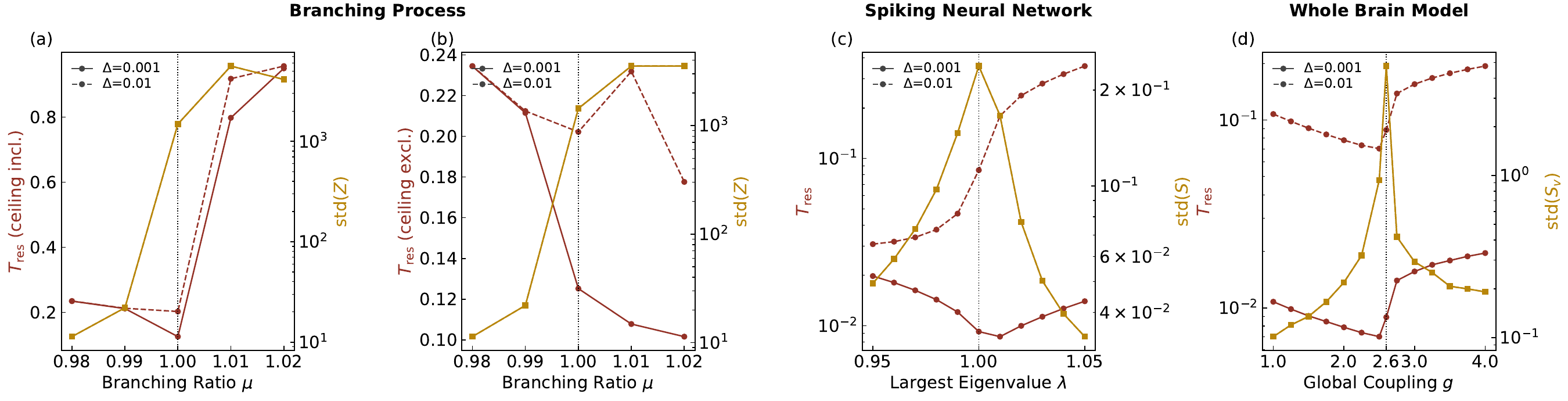}
    \caption{Residence time $T_{\text{res}}$ spent near $\sigma=1$,
and the standard deviation of the average activity $\text{std}_\text{t}(X)$, where X denotes the activity variable $Z$, $S$, and $S_v$ for the branching process (a, b), spiking neural network (c), and whole-brain model (d), respectively. Left and right $y$-axes correspond to $T_{\text{res}}$ and $\text{std}_\text{t}(X)$ respectively. Line styles (solid, dashed) distinguish the bin size values $\Delta$ in each panel, which are used to define $\sigma \approx 1$, i.e.\ $\sigma \in [1-\Delta, 1+\Delta]$. (a) All samples are included. (b) Samples near the population ceiling ($Z(t) > 0.9\cdot Z_m$) are excluded. Vertical dotted lines indicate the critical point $\mu_c=1$ in the branching process, $\lambda_c=1$ in the spiking neural network, and $g_c \approx 2.6$ in the whole-brain model (AAL90 parcellation). See Table~\ref{tab:control_param_summary} for the definition of the control parameters $\mu$, $\lambda$, and $g$.}
\label{fig:fig_0}
\end{figure*}

\section{Theoretical Background} 
Fisher information can be defined as the amount of information that an observable carries about the parameters that govern its probability distribution. %
By thinking of probability distributions as defining a manifold over their underlying parameters (e.g.\ mean and standard deviation for Normal distributions)~\cite{amari2016information}, one can consider the FIM as a distance measure, namely a metric on this statistical space~\cite{ruppeiner1979thermodynamics, janyszek1989riemannian, PhysRevE.51.1006,ruppeiner1995riemannian, kim2021information}. More formally, consider a system whose state is described by a random variable $X$ and a set of control parameters $\theta$. For a probability distribution function $P(X|\theta)$, FIM is defined as~\cite{amari2000methods, amari2016information}:
\begin{equation}
    G_{\theta_i \theta_j}=\int p(X|\theta)\frac{\partial \ln p(X|\theta)}{\partial \theta_i}\frac{\partial \ln p(X|\theta)}{\partial \theta_j} \mathrm{d}X \,,
    \label{Eq:fim_definition}
\end{equation}
which measures the amount of information the random variable $X$ carries about parameters $\theta$. The metric thus defines a dimensionless statistical distance ${ds^2 = \sum_{i,j} G_{\theta_i \theta_j} d \theta_i d \theta_j}$, providing a coordinate-independent measure of the distinguishability between nearby distributions~\cite{machta2013parameter, crooks2007measuring}.

Beyond its information-geometric origins, the FIM provides a natural link to statistical mechanics. In fact, for systems described by a Gibbs ensemble, FIM is equivalent to the thermodynamic metric tensor, effectively being the curvature of free entropy~\cite{PhysRevE.51.1006, brody2003information, janke2004information, crooks2007measuring,  crosato2018thermodynamics}. 
Within this framework, FIM serves as a generalized susceptibility that quantifies the response of the system to parameter variations~\cite{hidalgo2014information}, diverging in infinite systems or peaking in finite-size systems at criticality. Extensive previous work has successfully applied FIM to understand the onset of phase transitions in equilibrium and non-equilibrium systems~\cite{crosato2018thermodynamics, PhysRevE.91.062143, PhysRevE.93.023301,har2016information,Prokopenko2025}, inspiring us to use FIM to study the non-equilibrium burst activity in different systems operating near criticality. 

\section{Results}
Our analysis focuses on the behavior of three synthetic models that exhibit critical dynamics: a standard branching process, a spiking neural network~\cite{li2020tuning}, and a whole-brain model for BOLD human brain activity~\cite{deco2013resting, deco2014local, herzog2024neural}. 

Before turning to FIM, we first investigate the role of the \textit{observed} branching ratio $\sigma$ in the dynamics of these models, analyzing its behavior near the critical regime. These results will help interpret the FIM findings that follow.
Then, we evaluate FIM across these systems, first computing FIM with respect to each model's theoretical control parameter, which directly governs the distance from criticality, and then using the observed branching ratio (see Table~\ref{tab:control_param_summary} for a summary of the models' parameters).

Since the underlying distributions are not analytically tractable, in all three cases we estimate FIM from simulated data: the probability distribution $P(X|\theta)$ is first estimated non-parametrically, and the parameter interval is optimized to ensure that the estimated FIM remains within error tolerance~\cite{PhysRevE.93.023301,kinney2014estimation, bialek1996field}. We refer to the Supplementary Material (SM) for additional details.

\begin{figure*}[ht]
    \includegraphics[width=\linewidth]{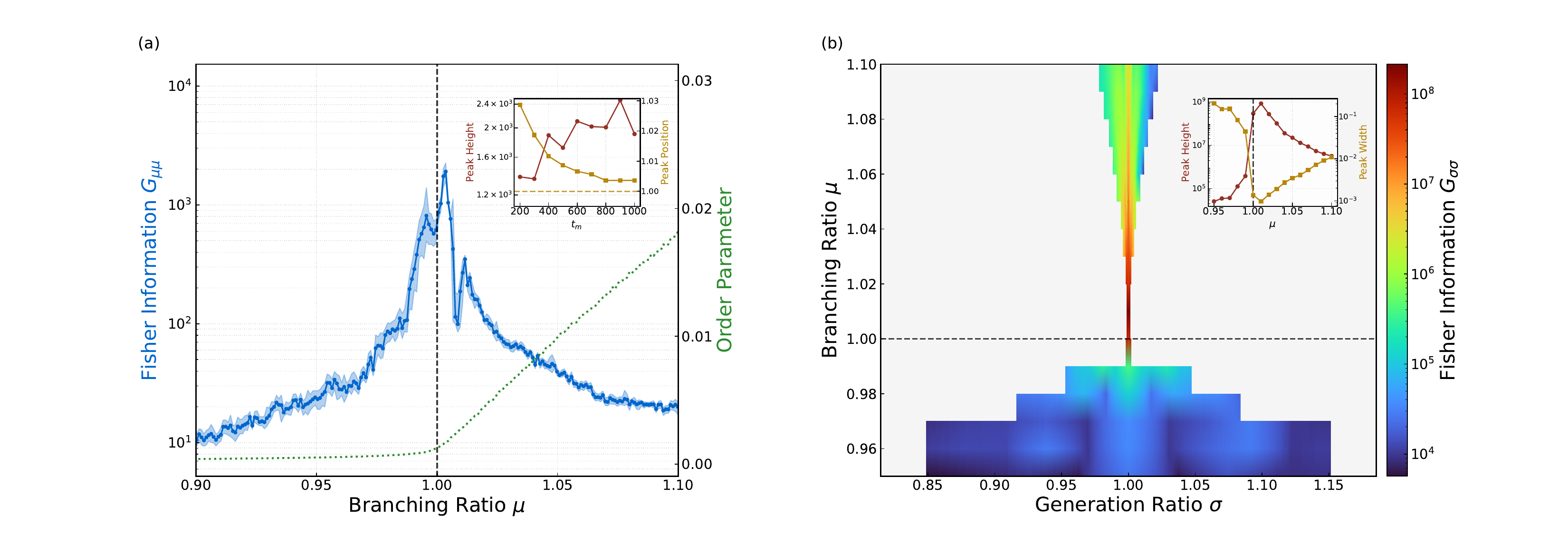}
    \caption{\textbf{Branching Process}. (a) Fisher information of the average branching ratio $G_{\mu\mu}$ for a system size of $t_m=1000$ shows a pronounced peak near the critical point $\mu_c=1$. The dotted line indicates the order parameter $t_\infty$. Inset: dependence of the peak height and peak position on system size $t_m$; the peak height increases with $t_m$, while the peak position converges to $\mu_c=1$ for large $t_m$. (b) Fisher information of the generation ratio $G_{\sigma\sigma}$ is shown as a heat map in the $(\mu,\sigma)$ plane. The Fisher information peaks at $\sigma=1$ throughout the critical regime, with the peak becoming sharpest and narrowest near $\mu_c=1$, as shown in the inset. The light gray region marks values outside the sampled range; only a finite set of $\sigma$ values close to $\sigma=1$ is evaluated. The peak width is quantified by the second central moment of the Fisher information, defined as $\sqrt{\sum_i w_i (\sigma_i-\sigma_{\mathrm{peak}})^2}$ with normalized weights $w_i \propto G_{\sigma \sigma}(\sigma_i)$~\cite{papoulis2002probability}. In both panels, Fisher information is estimated from bootstrap-resampled ensembles: from the $N_{\mathrm{esti}}$ raw samples, bootstrap resampling is performed to reconstruct the probability distribution function. In (a), the solid curve shows the bootstrap mean, with the shaded region being the 5th–95th percentile range. In (b), the heat map indicates the bootstrap-mean Fisher information. For algorithm details, see the Algorithm section in SM. The number of bootstrap $N_{\text{bootstrap}}$ is fixed at $32$ throughout the paper.}
    \label{fig:fim_branching_process}
\end{figure*}

\subsection{Observed Branching Ratio Residence Times}\label{branching_ratio_residence_time}
The focus of SOC on avalanche activity has motivated studies of the branching ratio as a way to monitor critical behavior. For a mathematical branching process, the branching ratio $\mu$ is defined in terms of the probabilities $p_k$ that a node produces $k$ offspring~\cite{harris1963theory, athreya2012branching}:
\begin{equation}
\mu=\sum_kkp_k \,,
\label{def_mu}
\end{equation}
This parameter indicates whether the activity, on average, grows ($\mu>1$), decays ($\mu<1$), or remains balanced ($\mu=1$), hence determining the supercritical, subcritical, and critical regimes of the system, respectively. Phenomenologically, one can consider the \textit{observed} branching ratio $\sigma(t)$ as the ratio of the number of offspring $Z$ at two consecutive time points:
\begin{equation} 
    \sigma(t) =  \frac{Z(t+1)}{Z(t)}\,,
    \label{def_sigma}
\end{equation}
Accordingly, $\sigma(t)$ quantifies the instantaneous growth or decay of the system's activity, and an analogous quantity can be defined for any system evolving over time by substituting the corresponding population-level observable for $Z(t)$ in Eq.~\eqref{def_sigma}. Thus, a natural question is whether the observed branching ratio $\sigma$ can provide a reliable empirical measure of criticality even outside the branching process. 

For a range of systems, including precipitation~\cite{peters2006critical}, neuronal activity~\cite{tagliazucchi2012criticality}, and the forest-fire model~\cite{jensen2021critical, palmieri2018emergence}, it has been reported that the system spends most of its time in the vicinity of the critical state. This behaviour can be quantified through the \textit{residence time}, i.e.\ the fraction of time that a given observable, tracked at each timepoint, spends within a small interval around a specified value. For an observable tracking the system's proximity to criticality, this implies that the residence time of the observable's critical value peaks when the system itself is globally critical, as measured by its control parameter.
However, we do not find this to be the case in the systems considered. Specifically, we examine the residence time of $\sigma$ together with the temporal standard deviation of the activity $\text{std}_\text{t}(\cdot)$, as a function of the relevant control parameter for all three models mentioned above (Fig.~\ref{fig:fig_0}).

The two neural systems exhibit broadly consistent behavior: the residence time displays a minimum at the critical point, while the standard deviation of the activity peaks, reflecting broad fluctuations near criticality. However, the branching process shows qualitatively different behavior, with a monotonic trend w.r.t.\ the model's control parameter, mainly related to difficulties handling branching processes that never terminate. 

Taken together, in all cases, the observed branching ratio does not exhibit a peak around $\sigma=1$. Hence, despite $\sigma$ being closely tied to the system's degree of criticality, controlled by $\mu$, the proximity to the critical regime cannot be directly inferred from the statistics of the observed branching ratio. As we demonstrate below, this contrasts with the FIM of $\sigma$, which displays a clear peak near the critical region.

\begin{figure*}[ht]
    \includegraphics[width=\linewidth]{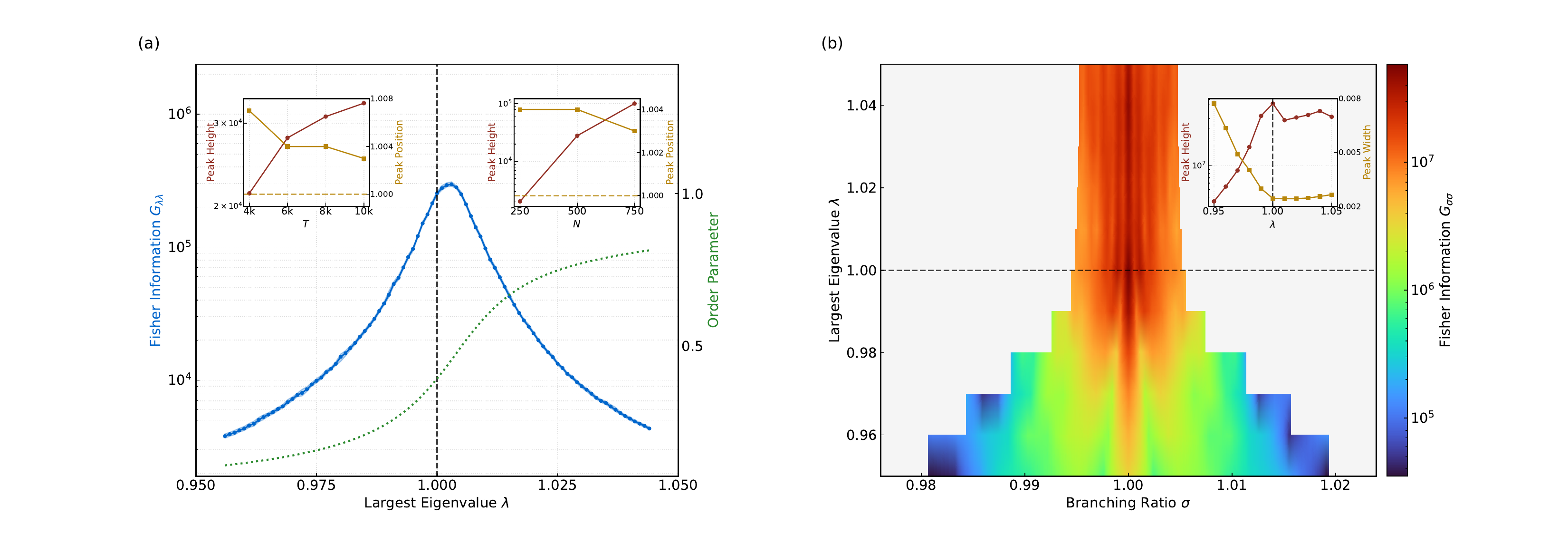}
    \caption{\textbf{Spiking Neural Network}. (a) Fisher information of the largest eigenvalue, $G_{\lambda \lambda}$, for system size $N=1000$ shows a pronounced peak near the critical point $\lambda_c=1$. The dotted line indicates the order parameter $S$.   
    Insets: dependence of the peak height and peak position on the spatial and temporal system size $N, T$, respectively; the peak height increases with $N, T$, while the peak position converges to $\lambda_c=1$ for large $N, T$. (b) Fisher information of the observed branching ratio, $G_{\sigma\sigma}$, shown as a heat map in the $(\lambda,\sigma)$ plane. The Fisher information peaks at $\sigma=1$ throughout the critical regime, with the peak becoming sharpest and narrowest near $\lambda_c=1$, as shown in the inset. The light gray region marks values outside the sampled range; only a finite set of $\sigma$ values close to $\sigma=1$ is evaluated.}
    \label{fig:fim_spiking_brain_model}
\end{figure*}

\subsection{Branching Process}
\label{Branch_proc}
We begin by considering the Galton-Watson branching process as a test bed to validate our approach for the estimation of FIM~\cite{athreya2012branching, harris1963theory}. Since this is an extremely simplified and controlled scenario, it also helps us to understand the interplay between the \textit{observed} branching ratio $\sigma(t)$ (see Eq. (\ref{def_sigma})) and the underlying parameters. Within this model, the control parameter is given by the \textit{true mathematical} average branching ratio $\mu$ (see Eq. (\ref{def_mu})), corresponding to the  average number of offspring each node produces computed from the branching probabilities~\cite{athreya2012branching, harris1963theory}. Calculating the FIM $G_{\mu\mu}$ for such a parameter, we obtain a sharp peak at $\mu = 1$, as expected (Fig.~\ref{fig:fim_branching_process}(a)). In particular, the peak height increases with the maximum number of generations ($t_m$) while the peak position converges to $\mu_c = 1$, confirming the robustness and sensitivity of FIM. 

We then consider FIM of the dynamical observable $\sigma(t)$, scanning a range of $\mu$ values to examine the behavior of FIM near the critical regime. Results show that $G_{\sigma\sigma}$ exhibits peaks at $\sigma = 1$ across all values of $\mu$ explored, ranging from subcritical to supercritical regimes (Fig.~\ref{fig:fim_branching_process}(b)). Also, the structure of this peak varies systematically with $\mu$, exhibiting the highest and the narrowest peaks in the region near $\mu =1$, as the system is maximally sensitive to deviations of $\sigma$ from unity. On the other hand, as $\mu$ departs from criticality, the peak height decreases while its width broadens, indicating a loss of sensitivity. 
Notably, the broadening is asymmetric around the critical point $\mu=1$, plausibly due to the asymmetry between branching processes generated when $\mu<1$ and $\mu>1$. In fact, in the subcritical regime $\mu<1$, the average generation size exponentially decreases, and hence the generation size $Z(t)$ tends to be small. On the other hand, in the supercritical regime, the average generation size grows exponentially, and there is a non-zero probability that a realization of the process never stops.

\subsection{Spiking Neural Network}
\label{spiking_brain}
We next consider a spiking neural network composed of $N$ excitatory and inhibitory neurons~\cite{larremore2014inhibition, li2020tuning}, in which each node is either quiescent or active. A quiescent node becomes active at the next time step with a probability determined by the weighted sum of inputs from its currently active neighbors, where excitatory neighbors increase this probability, and inhibitory neighbors decrease it. 
The critical dynamics of this network is governed by the largest eigenvalue $\lambda$ of the connectivity matrix (see Sec.~\ref{sec:method_spiking} for more details). Although this model lacks the biophysical complexity of a real biological circuit, it incorporates both excitatory and inhibitory populations and exhibits a genuine phase transition at $\lambda_c = 1$~\cite{larremore2011predicting}.

Hence, we first consider the largest eigenvalue $\lambda$ as the control parameter for FIM, observing that $G_{\lambda\lambda}$ exhibits a sharp peak around the true critical point $\lambda_c = 1$~(Fig.~\ref{fig:fim_spiking_brain_model}(a)). As in the case of the branching process, the peak height increases with both the spatial ($N$) and temporal ($T$) system size, consistent with a divergence of susceptibility in the infinite system limit. 

We then calculate $G_{\sigma\sigma}$, i.e.\ the FIM of the observed branching ratio $\sigma$ defined on the network's activity ratio (see Table~\ref{tab:control_param_summary}). Similarly to the previous model, $G_{\sigma\sigma}$ exhibits the highest and narrowest peak at $\sigma=1$ and $\lambda=1$, which then broadens progressively as the system moves from criticality to subcriticality (Fig.~\ref{fig:fim_spiking_brain_model}(b)). However, the supercritical regime is also characterized by a peak that remains largely invariant of $\lambda$. This unexpected behavior is plausibly a direct consequence of the inhibitory feedback that prevents activity from saturating, bounding the average activity $S(t)$ regardless of $\lambda$~\cite{larremore2014inhibition}. Finally, finite-size analysis of $G_{\sigma\sigma}$ at $\lambda=1$ reveals that as $N$ grows, the peak increases and narrows, providing further evidence that $\sigma=1$ is a genuine critical point for FIM in the infinite system limit~(see Fig.~2 in SM).

\begin{figure*}[ht]
    \includegraphics[width=\linewidth]{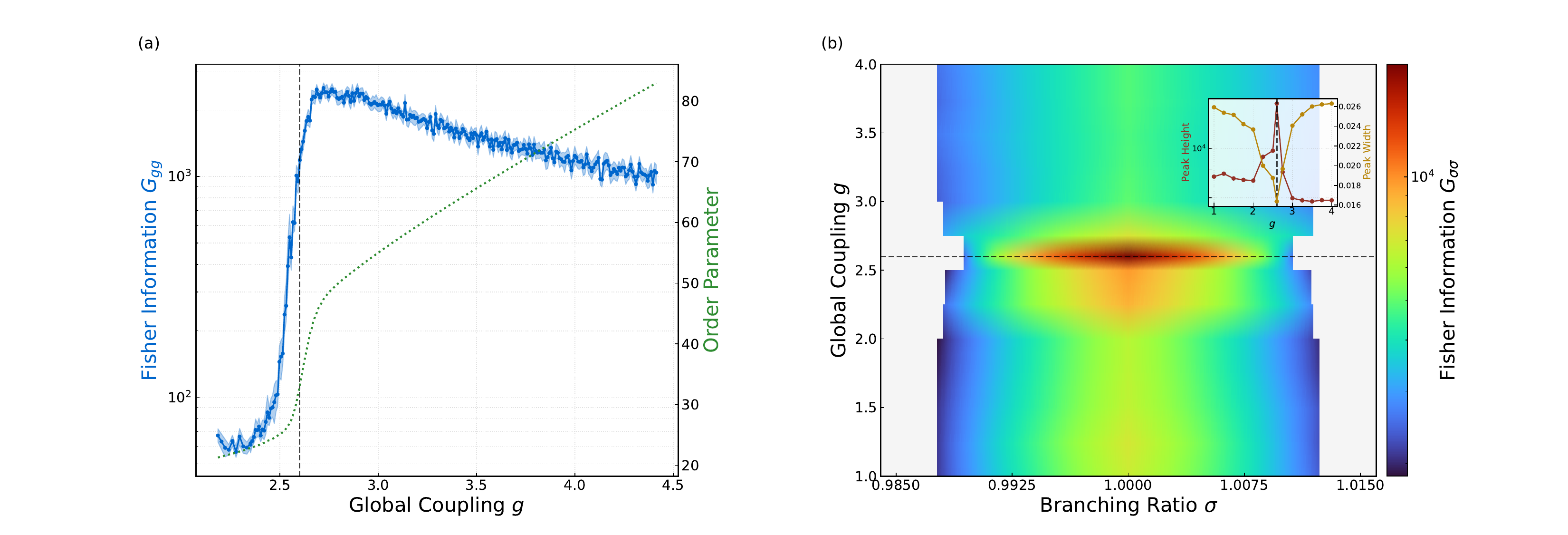}
    \caption{\textbf{Whole-Brain Model}. (a) Fisher information of the global coupling $G_{gg}$ for AAL90 parcellation with 90 brain regions, showing a pronounced peak near $g_c \approx2.6$. The dotted line indicates the order parameter $S_v$. (b) Fisher information of the observed branching ratio $G_{\sigma\sigma}$ is shown as a heat map in the $(g,\sigma)$ plane. The Fisher information peaks at $\sigma=1$ throughout the critical regime, with the peak becoming sharpest and narrowest near $g_c\approx2.6$, as shown in the inset. The light gray region marks values outside the sampled range; only a finite set of $\sigma$ values close to $\sigma=1$ is evaluated. For the same analysis of other parcellations (Schaefer100, Schaefer1000), see SM Fig.~5, 6.
    }
    \label{fig:fim_whole_brain_model_c90}
\end{figure*}

\subsection{Whole-Brain Model}
\label{sec:whole_brain}
We finally apply FIM to the Dynamic Mean Field (DMF) whole-brain model~\cite{deco2013resting, deco2014local, herzog2024neural}, a biophysically grounded model whose regions are coupled via an empirical structural connectome derived from diffusion tensor imaging (DTI)~\cite{van2012human}~(see Sec.~\ref{sec:method_dmf} for details). This model represents the most realistic setting considered here, incorporating structural heterogeneity, excitation-inhibition balance constraints, and generating reliable BOLD signals. This enables us to test whether the methodology can identify criticality in a system closely constrained by human neuroimaging data.

We first consider FIM of the global coupling $g$, which governs the model's proximity to its optimal working point. Given the spatial heterogeneity of the model, we account for the structural diversity by defining the global activity as a weighted average firing rate, where the weights are given by the largest eigenvector of the structural connectivity. $G_{gg}$ displays a pronounced peak near $g_c \approx 2.6$, agreeing with the sharp transition of the order parameter
(Fig.~\ref{fig:fim_whole_brain_model_c90}(a)). 

Then, we consider FIM of the observed branching ratio $\sigma(t)$: remarkably, $G_{\sigma \sigma}$ exhibits the same geometric signature as in the previous simpler models: a peak at $\sigma = 1$ that is sharpest and narrowest near $g_c$ and that broadens progressively as the system deviates from criticality (Fig.~\ref{fig:fim_whole_brain_model_c90}(b)). The results are robust to the choice of threshold within a broad range~(Fig.~4 in SM). 
Thus, the robustness of these findings in such a biophysically detailed model demonstrates that FIM of the observed branching ratio can locate signatures of critical behavior even in complex, heterogeneous dynamics. In SM, we further show that FIM peak height grows with system size, consistent with the finite-size effects near criticality.

\section{Discussions}
In this paper, we presented the Fisher Information Metric (FIM) as an information-geometric susceptibility for quantifying proximity to criticality in models of neuronal dynamics. The FIM of the theoretical control parameter accurately identifies the critical region in all cases. However, since the theoretical control parameter is often not empirically accessible, we proposed a data-driven approach that leverages the observed branching ratio $\sigma$. We have demonstrated that both the theoretical and the phenomenological FIM peak sharply at the critical point, with the peak becoming higher and narrower as the system size increases. However, we remark that these two approaches capture fundamentally and conceptually different aspects of the dynamics, which we discuss below.

\subsection{Global versus instantaneous criticality}

Despite the similar results, our twofold approach characterizes profoundly distinct aspects of the systems' criticality. 
On the one hand, the FIM of the theoretical control parameter identifies the global operating regime of the system and its critical point. By contrast, the FIM with respect to the observed branching ratio $\sigma$ characterizes the statistical sensitivity of the instantaneous dynamical state. In other words, it describes the system's sensitivity to its parameters when the system's control parameter is fixed, as different dynamical regimes unfold over time. The consistent appearance of a peak at $\sigma=1$ is a direct consequence of this approach, indicating that self-replicating activity corresponds to a temporal transient of criticality regardless of whether the system is globally critical. Moreover, the dependence of the peak height and width on the theoretical control parameter provides a measure of how close the system operates to the global critical point. In this sense, the theoretical control parameter determines the global distance from criticality, whereas the observed branching ratio quantifies how closely the instantaneous dynamics resembles criticality within that operating regime. In the two network models considered here, this distinction also reflects a separation of timescales: the theoretical control parameter is determined by the underlying connectivity and changes only through structural modifications, whereas the observed branching ratio fluctuates in time on a fixed network. This may suggest the underlying mechanism by which biological networks may dynamically regulate their proximity to criticality without requiring structural rewiring on short timescales~\cite{parks2024nonoscillatory,ma2019cortical,shew2015adaptation,fontenele2019criticality,xu2024sleep}.

\subsection{FIM as a robust non-parametric measure of criticality}
As discussed in Sec.~\ref{intro}, traditional empirical estimators of neuronal criticality are complicated by methodological choices and modeling assumptions that can bias the inferred distance to criticality~\cite{dehghani2012avalanche, bedard2009macroscopic, 
touboul2010can, klaus2011statistical, PRXLife.3.013013, Touboul_2017}. The FIM approach proposed here sidesteps these issues, operating directly on continuous activity signals with no need to define discrete avalanche events. Furthermore, as a generalized susceptibility, FIM is expected to diverge at the critical point in the infinite system size limit and obeys standard finite-size scaling. As we demonstrated in both the branching process and the spiking model, peak height grows with system size while peak position converges to the true critical point, allowing finite-size effects to be systematically investigated, though a comprehensive scaling analysis is too numerically demanding to be included in the present study.

\subsection{Network heterogeneity and distance to criticality}
Network heterogeneity fundamentally complicates the notion of ``distance to criticality''. In the whole-brain model considered here, we weight regional activity by the largest eigenvector of the structural connectivity, to account for substantial heterogeneity across brain regions. This choice reflects a broader issue in heterogeneous networks that different regions can sit at different effective distances from their local critical points~\cite{rusch2025influence, larremore2011predicting}, which can further give rise to extended critical-like regimes such as Griffiths phases~\cite{moretti2013griffiths}. Therefore, while a global FIM provides a useful characterization of the collective operating regime, spatially resolved measures of proximity to criticality represent an important direction for future work.

\subsection{Limitations and future work}
The analysis of criticality in terms of the FIM, however, faces practical limitations. The currently adopted non-parametric estimation of FIM is data-intensive in both spatial and temporal dimensions. This requires empirical data to be collected at fine-grained spatial resolution and over long observation windows to reliably estimate the underlying probability distributions. Beyond data requirements, additional challenges arise from the methodological choices inherent in FIM estimation: the binarization threshold, the optimal bin width introduced by the adaptive binning strategy, and relevant hyperparameters in probability density estimators are all free parameters whose fine-tuning can influence the resulting FIM estimates. However, although in principle these effects can strongly bias FIM, in practice we noticed that the results are robust to various choices of the aforementioned parameters (see SM).
A more fundamental requirement is that the system must have reached stationarity, and deviations from it can bias the estimated distributions and therefore FIM. Extending the framework to empirical neural recordings is thus a promising avenue for future research, as assessing and ensuring stationarity in experimental data remains methodologically challenging.

\subsection{Conclusions}
In summary, we found the Fisher Information Metric to provide a powerful and practically accessible framework for characterizing criticality in synthetic neural systems. The FIM of an activity-dependent observable serves simultaneously as a detector of criticality and a quantitative measure of proximity to it, without requiring access to the microscopic control parameter or the definition of discrete avalanche events. This model-agnostic property makes FIM a promising tool for measuring distance to criticality from empirical neural recordings (e.g.\ from multi-electrode array recordings, calcium imaging, or fMRI time series), where true control parameters are inherently inaccessible and only measurements of instantaneous activity are available.

\section{Methods}

\subsection{Model Definitions and choice of parameters and variables of FIM}

\subsubsection{Branching process}
\label{sec:method_branching}
The Galton-Watson branching process is a discrete-time stochastic model used to describe population evolution~\cite{athreya2012branching, harris1963theory}. Starting from a single node $Z(t=0)=1$, where $Z(t)$ denotes the population size at generation $t$, each node $i$ independently produces a random number of offspring $x_i$, which is drawn from a Poisson distribution with mean $\mu$, such that the total population evolves according to:
\begin{equation}
    Z(t+1) = \sum_{i=1}^{Z(t)} x_i, \quad x_i \sim \text{Poisson}(\mu),
\end{equation}
where $\mu =\sum_{k=1}^\infty kp_k$ is the mathematical true average branching ratio, i.e.\ the expected number of offspring per individual at each timestep, with $p_k$ denoting the probability that a node produces $k$ offspring. The system exhibits a critical point at $\mu_c = 1$, meaning that both the distribution of total progeny size and of extinction times follow power-laws with exponents $3/2$ and $2$, respectively. For $\mu<1$, these distributions decay exponentially at large progeny size and extinction time. For $\mu>1$, there is a non-zero probability that a process never stops, complicates the analysis of the supercritical regime~\cite{jensen2022complexity, harris1963theory}. 

We first consider FIM of the theoretical control parameter by choosing the stochastic variable to be the rescaled number of surviving generations $t/t_m$ and the parameter to be $\mu$, with $t_m$ being the maximum number of generations in the simulations. Its average $t_\infty \equiv \lim_{t_m\rightarrow\infty} \langle t\rangle/t_m$ can be considered an order parameter since we have $t_\infty = 0$ for $\mu<1$ and $t_\infty>0$ for $\mu>1$. To ensure the estimated distributions reflect intrinsic dynamics rather than finite-size artifacts, we retain only finite branching trees and discard samples where $Z(t) > 0.9\cdot Z_m$, with $Z_m$ being the maximum allowed population in generation $t$. 

We then turn to FIM of the generation ratio $G_{\sigma \sigma}$, with $\sigma = Z(t+1)/Z(t)$ as the parameter and $Z(t)$ as the variable. When estimating $G_{\sigma\sigma}$ non-parametrically, we employ an adaptive binning strategy: digitizing $\sigma$ into bins centered at $\sigma=1$ and expanding outward, where the bin width is minimized on the condition that each bin contains at least $N_\text{esti}$ samples, balancing fine resolution with reliable statistics. The same binning strategy is applied in the $G_{\sigma \sigma}$ calculation throughout the three models.

It is worth noting that $\sigma(t)$ and $\mu$ are, in general, distinct quantities, and the time-averaged $\sigma(t)$ obtained from a single realization need not converge to the true branching ratio $\mu$. This reflects a more fundamental difficulty in estimating $\mu$ from finite statistics: the average of $\sigma(t)$ over a given branching tree depends on the duration of that specific realization, with short-lived trees yielding systematically smaller averages than long-lived ones, while the longest-lived realizations are themselves difficult to sample adequately in finite simulations. An additional complication comes from $Z(t)$ being an integer and, for small-size generations, i.e.\ $Z(t)$ and $Z(t+1)$ equal to just a few nodes, the ratio $\sigma(t)$ does not vary continuously about 1. 
In Sec.~\ref{branching_ratio_residence_time}, we show that this gives rise to the anomalous behavior of the residence time of $\sigma(t)$ in the branching process, contrasting the other two systems considered.

\subsubsection{Spiking neural network}
\label{sec:method_spiking}
We adopt the excitable network model introduced in ~\cite{larremore2014inhibition, li2020tuning}, consisting of $N$ nodes that are either quiescent ($s_n(t)=0$) or active ($s_n(t)=1$). The network is described by an $N\times N$ connectivity matrix $A$, where each node connects to any other with probability $p$. The state of node $n$ is then updated according to the rule:
\begin{equation}
    s_n(t+1) = 1 \quad \text{with probability} \quad p=f\!\left(\sum_{m=1}^{N} A_{nm}\, s_m(t)\right),
\end{equation}
where $A_{nm}$ is the connection strength from node $m$ to $n$, and $f(\cdot)$ is a piecewise-linear transfer function. The connection strengths of excitatory nodes are drawn from the uniform distribution $\mathcal{U}[0, w]$, while those of inhibitory nodes (with a fraction of $\alpha=0.2$) are drawn from another uniform distribution $\mathcal{U}[0, -gw]$, where $g$ is the relative strength and $w$ is the average synaptic weight. A probability of $p_{\text{ext}}= 0.005/N$ is introduced to describe external inputs.

The network can be understood as a correlated branching process where the activity propagation is governed by the largest eigenvalue $\lambda$ of the connectivity matrix $A$. Specifically, $\lambda$ serves as an ideal control parameter to tune the network towards or away from criticality. The system undergoes a second-order phase transition at $\lambda_c=1$, separating a quiescent phase ($\lambda<1$) from an overactive phase ($\lambda>1$)~\cite{larremore2014inhibition, larremore2011predicting, larremore2011effects, PhysRevE.85.066131, PhysRevE.86.021909}. 

We therefore choose $\lambda$ as the control parameter for $G_{\lambda\lambda}$, with the spatio-temporal average firing rate $S$ as the variable, $S = \sum_t S(t)/T,~S(t) = \sum_n s_n(t)/N$. For the FIM of the dynamical observable, we define the branching ratio as the parameter and the instantaneous firing rate $S(t)$ as the variable, computing Fisher information $G_{\sigma\sigma}$. Unlike for the whole-brain model below, the uniform spatial average is natural here because the network has i.i.d.\ connection weights with nodes of the same type being statistically equivalent. Furthermore, because inhibitory nodes introduce negative weights into the connectivity matrix $A$, the Perron--Frobenius theorem does not apply, and the components of the leading eigenvector of $A$ are not guaranteed to be sign-definite, precluding its use as a spatial weight, which we instead introduce in the following model.

\subsubsection{Whole-brain model}
\label{sec:method_dmf}
Dynamic Mean Field (DMF) model is a biophysically grounded whole-brain model derived by applying a mean-field reduction to single-neuron dynamics~\cite{deco2013resting, deco2014local}. Here, each of the $N$ brain regions is modeled as two interacting neural populations, one excitatory (E) and one inhibitory (I), whose local dynamics determine the mean firing rate of the macroscopic region. Regions are coupled via a structural connectivity matrix $C$, empirically derived from diffusion tensor imaging (DTI), where the global coupling parameter $g$ scales the strength of interregional excitatory inputs. We use three empirical connectomes $C$, AAL90, Schaefer100, and Schaefer1000~\cite{schaefer2018local, herzog2024neural}, with AAL90 and Schaefer1000 connectome rescaled by a factor of 0.2 following previous approaches in the literature~\cite{deco2013resting, herzog2024neural}. The simulated firing activity of each excitatory pool is then transformed into blood-oxygen-level-dependent (BOLD) signals~\cite{stephan2007comparing}, enabling direct comparison with empirical fMRI recordings.
 
To stabilize the firing rates of excitatory populations within a neurobiologically plausible range, a feedback inhibition control~(FIC) parameter $J_n$ is assigned to each region, denoting the local connection from inhibitory to excitatory population. Following Ref.~\cite{deco2014local}, $J_n^{\text{opt}}$ is determined by a recursive adaptation algorithm to compensate for the excess of long-range excitation. This procedure has then been more efficiently implemented in FastDMF, , a computationally efficient implementation of the DMF model that replaces the recursive FIC optimization with a connectome-dependent analytical solution and substantially reduces the computational cost of model fitting~\cite{herzog2024neural}.
Specifically, FastDMF proposes a first-order linear solution of the FIC optimization, $J_n^{\text{opt}} = a g \beta_n + 1$,  proportional to both $g$ and the local connectivity strength $\beta_n = \sum_p C_{np}$, encoding the connectome-dependent E/I balance. Moreover, FastDMF leverages a Bayesian optimization algorithm to fit the model to empirical fMRI data with fewer iterations.
 
The whole-brain model undergoes a dynamical phase transition as $g$ is varied. In the subcritical regime, regional dynamics are dominated by local fixed points, producing low-amplitude, weakly correlated BOLD signals. Near the critical coupling $g_c$, the network exhibits a transition toward a more complex, high-variance regime where long-range correlations emerge, and the model best reproduces empirical functional connectivity~\cite{deco2013resting, deco2014local}. The global coupling thus plays a role analogous to $\mu$ in the branching process and $\lambda$ in the spiking model (see Table~\ref{tab:control_param_summary}).

We first calculate $G_{gg}$. Unlike the spiking model, the empirical DTI connectome exhibits strong structural heterogeneity, with hub regions dominating collective dynamics with disproportionately large node strength $\beta_n$. A uniform spatial average would therefore dilute the signal from these hub regions. We instead define the variable as the eigenvector-weighted firing rate
\begin{equation}
    S_v(t) = \sum_n v_n r_n(t)
\end{equation}
where $r_n(t)$ is the continuous firing rate of region $n$ and $v_n$ is the normalized $n$-th component of the largest eigenvector of the connectome $C$. This identifies regions most involved in the dominant collective mode, analogous to how the largest eigenvalue $\lambda$ of $A$ controls the critical dynamics in the spiking model~\cite{larremore2011effects, larremore2011predicting}. The scalar observable is 
then $S_v = \max_{\text{t}} S_v(t)$.

We finally consider the FIM $G_{\sigma \sigma}$, with the variable being the continuous weighted firing rate $S_v(t)$. The observed branching ratio $\sigma$ is calculated based on binarized firing rates. The binarization converts the continuous firing rate of each region into a binary active/inactive sequence $\hat{s}_n(t) \in \{0,1\}$: region $n$ is active at time $t$ if its firing rate exceeds the $p$-th percentile of its own firing rate distribution across all time steps, and inactive otherwise. Here we chose $p=50$, i.e.\ its median. The binary signal on the network scale is defined as the uniform spatial average $\hat{S}(t) = \sum_n \hat{s}_n(t)/N$, rather than the eigenvector-weighted average used for the continuous signal. The branching ratio is then defined as $\sigma(t) = \hat{S}(t+1)/\hat{S}(t)$. This approach to computing the observed branching ratio is motivated by the following considerations. The binarization of continuous signals avoids spurious extreme ratios when the continuous signal approaches zero, while preserving the information on activity growth and decay that is sufficient to identify the vicinity of $\sigma=1$. Then, per-region percentile thresholding effectively normalizes each region to the same scale, eliminating the inter-region heterogeneity in baseline firing rates that motivated eigenvector weighting. A uniform average over $\hat{s}_n(t)$ is therefore the natural choice. The resulting $\hat{S}(t)$ represents the instantaneous fraction of active units, analogous to the average network activity $S(t)$ in the spiking neural network, ensuring that the observed branching ratio is defined consistently across both models.

\section{Acknowledgements}
The authors acknowledge the High-Performance Computing (HPC) provided by the Imperial College Research Computing Service (DOI: 10.14469/hpc/2232). H.R. is funded by the Schmidt Sciences LLC. through its AI in Science Fellowship program.

\bibliographystyle{unsrt}
\bibliography{mybib}

\end{document}


\preprint{APS/123-QED}

\title{Supplementary Material}%

\author{Yuewei Du$^{1}$}
\author{Alberto Liardi$^{2,3}$}
\author{Hardik Rajpal$^{3,5,6}$}
\author{Henrik Jeldtoft Jensen$^{3,4}$}  \email{h.jensen@imperial.ac.uk}

\affiliation{$^{1}$Department of Physics, Imperial College London, London, United Kingdom}
\affiliation{$^{2}$Department of Computing, Imperial College London, London, United Kingdom}
\affiliation{$^{3}$Centre for Complexity Science, Imperial College London, London, United Kingdom}
\affiliation{$^{4}$Department of Mathematics, Imperial College London, London, United Kingdom}
\affiliation{$^{5}$I-X Centre for AI in Science, Imperial College London, London, United Kingdom}
\affiliation{$^{6}$Department of Electrical and Electronic Engineering, Imperial College London, London, United Kingdom}
\date{\today}%

\maketitle

\section{Algorithm}
\label{sec:FIM_estimation}
Since the underlying distributions are not analytically tractable in general, we estimate 
the FIM directly from empirical data in two steps. First, given $N_{\text{esti}}$ i.i.d.\ samples at 
each parameter point $\theta$, we estimate the probability distribution $P(X|\theta)$ 
nonparametrically using the Density Estimation using Field Theory (DEFT) 
method~\cite{kinney2014estimation}, for which a Python implementation is publicly 
available~\cite{deft}. Then, with $P(X|\theta)$ estimated at each parameter point, the 
FIM components are computed via finite differences:
\begin{equation}
\begin{aligned}
G_{\theta_{\mu} \theta_{\nu}}(\theta) \approx \int &\frac{p(X|\theta+\Delta\theta^\mu)-p(X|\theta-\Delta\theta^\mu)}{2\Delta\theta^\mu} \\
&\times \frac{p(X|\theta+\Delta\theta^\nu)-p(X|\theta-\Delta\theta^\nu)}{2\Delta\theta^\nu} \frac{dx}{p(X|\theta)}
\end{aligned}
\end{equation}
where $\Delta\theta^\mu$ is the finite-difference step size along the $\mu$-th parameter 
direction. This centered finite difference is used by default throughout the paper as it 
typically provides higher accuracy and better numerical stability. However, specific care 
should be taken when computing $G_{\sigma\sigma}$, where probability distributions at 
nearby parameter points differ substantially. To maintain numerical stability, we adopt the 
logarithmic definition of the Fisher information, and estimate $G_{\sigma\sigma}$ 
independently via forward and backward differences, taking their average as the final result.

The accuracy of the FIM estimate depends critically on the step size $\Delta\theta^\mu$. 
Following~\cite{PhysRevE.93.023301}, a dimensionless parameter 
$\epsilon^2 = \frac{2}{N_{\text{esti}}\, G_{\theta_{\mu} \theta_{\nu}}(\theta)\,\Delta\theta^\mu \Delta\theta^\nu}$ 
is introduced to quantify the numerical accuracy, where $N_{\text{esti}}$ is the number of samples. 
It has been shown that $\epsilon \sim 0.05$ yields tolerable errors~\cite{PhysRevE.93.023301}, 
and we therefore adopt $\epsilon^* = 0.05$ with a tolerance of $\pm 0.02$. 
The optimal $\Delta\theta^\mu$ is determined via a bisection method: starting from the 
smallest feasible step size, $\Delta\theta^\mu$ is iteratively adjusted---increased if 
$\epsilon > \epsilon^*$ and decreased otherwise---until $\epsilon$ falls within the 
tolerance band. All key algorithmic parameters used are summarized in Table~\ref{tab:FIM_params}.

\begin{table}[h]
\centering
\caption{Algorithm parameters for FIM estimation across all systems. 
$G$ denotes the number of grid points for DEFT density estimation, 
$N_\mathrm{esti}$ the number of samples used, and $\alpha$ the DEFT 
smoothness parameter (set to $\alpha=1$ throughout).}
\label{tab:FIM_params}
\begin{ruledtabular}
\begin{tabular}{llcc}
System & FIM & $G$ & $N_\mathrm{esti}$\\
\hline
\multirow{2}{*}{Branching process} 
  & $G_{\mu\mu}$           & 100 & 100{,}000 \\
  & $G_{\sigma\sigma}$     & 200 &  10{,}000 \\
\hline
\multirow{3}{*}{Spiking neural network}       
  & $G_{\lambda\lambda}$   & 100 &  10{,}000 \\
  & $G_{\sigma\sigma}$     &  50 &  10{,}000 \\
  & $G_{nn}$     &  100 &  100{,}000 \\
\hline
\multirow{2}{*}{Whole-brain model} 
  & $G_{gg}$               & 100 &  10{,}000 \\
  & $G_{\sigma\sigma}$     & 200 &  10{,}000 \\
\hline
\multirow{1}{*}{Real data}                
  & $G_{\sigma\sigma}$     &  50 &  10{,}000 \\
\end{tabular}
\end{ruledtabular}
\end{table}

\section{Branching Process}
\begin{figure}[!htbp]
    \includegraphics[width=\linewidth]{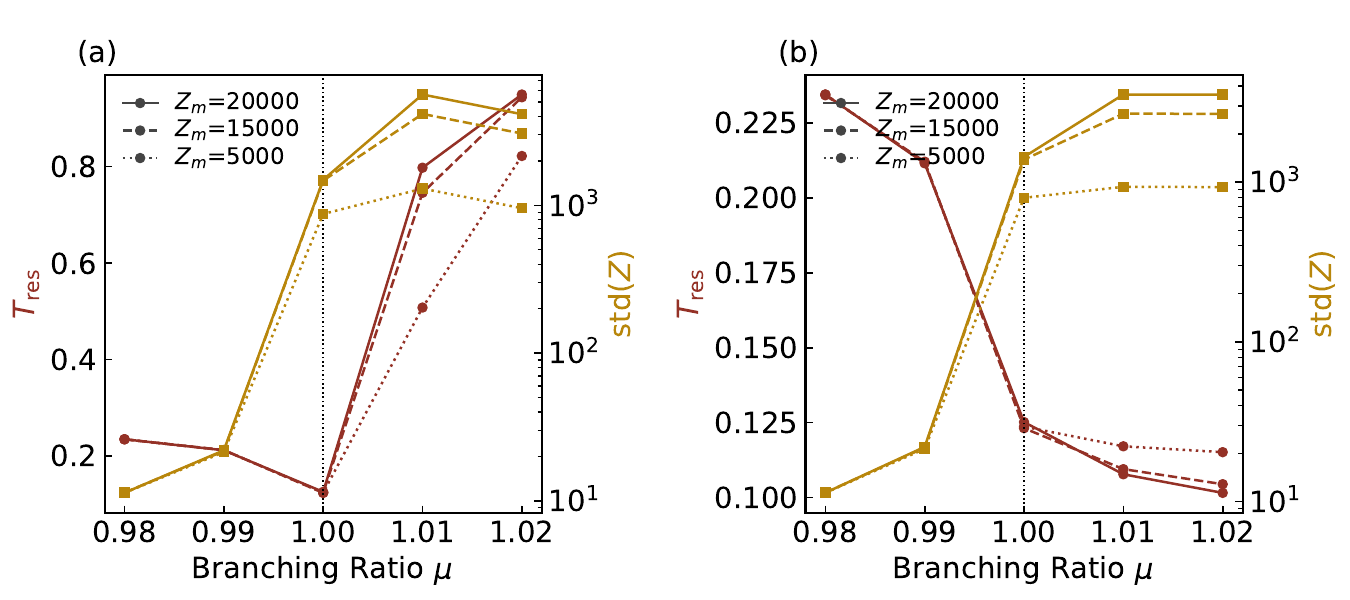}
    \caption{Residence time $T_{\text{res}}$ spent near $\sigma=1$, and the standard deviation of the average firing rate over the time series $\text{std}_\text{t}(Z)$ in the branching process simulated for a maximum of  $t_m=10000$ generations. 
    Left and right $y$-axes correspond to $T_{\text{res}}$ and $\text{std}_\text{t}(Z)$ respectively. Line styles (solid, dashed, dotted) distinguish the maximum allowed population $Z_m$ in numerical simulations in each panel. (a) All samples included. (b) Samples near the population ceiling, $Z(t) > 0.9 *Z_m$, removed. Vertical dotted lines mark the critical point $\mu_c = 1$.}
\label{fig:T_res_bs}
\end{figure}

Fig.~\ref{fig:T_res_bs} shows the residence time and temporal fluctuations in the branching process are not sensitive to the maximum allowed populations in a generation $Z_m$.

\section{Spiking neural network}
\begin{figure}[!htbp]
    \includegraphics[width=\linewidth]{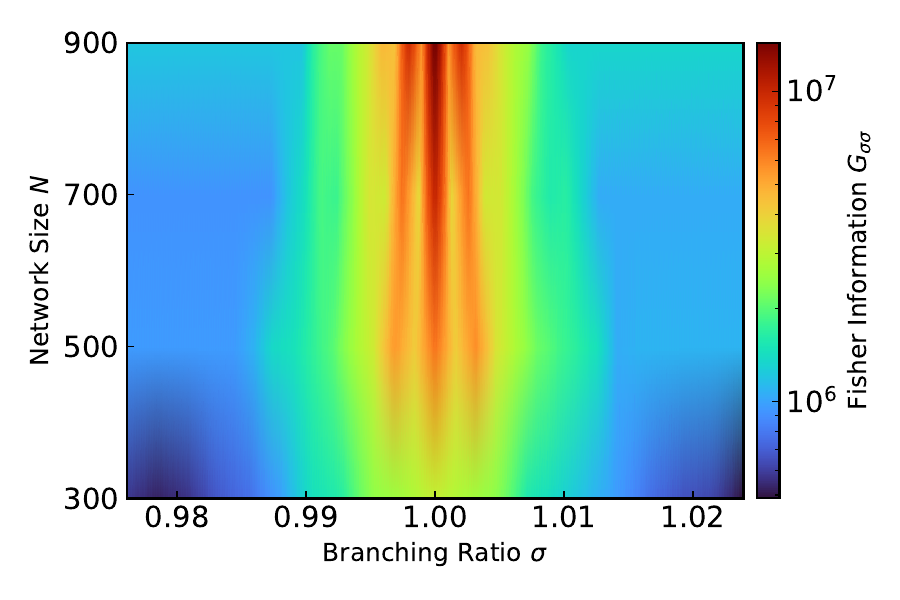}  
    \caption{Finite size effects of Fisher information $G_{\sigma \sigma}$ in spiking neural network. Increasing network size $N$ increases the peak height and decreases the peak width at $\sigma=1$.}
    \label{fig:fim_of_sigma_across_N}
\end{figure}

In Fig.~\ref{fig:fim_of_sigma_across_N} we perform the finite-size analysis of $G_{\sigma\sigma}$ at criticality $\lambda=1$, which reveals that as $N$ grows, the peak increases and narrows, providing further evidence that $\sigma=1$ is a genuine critical point in the infinite system size limit. 

\begin{figure*}[!htbp]
    \includegraphics[width=\linewidth]{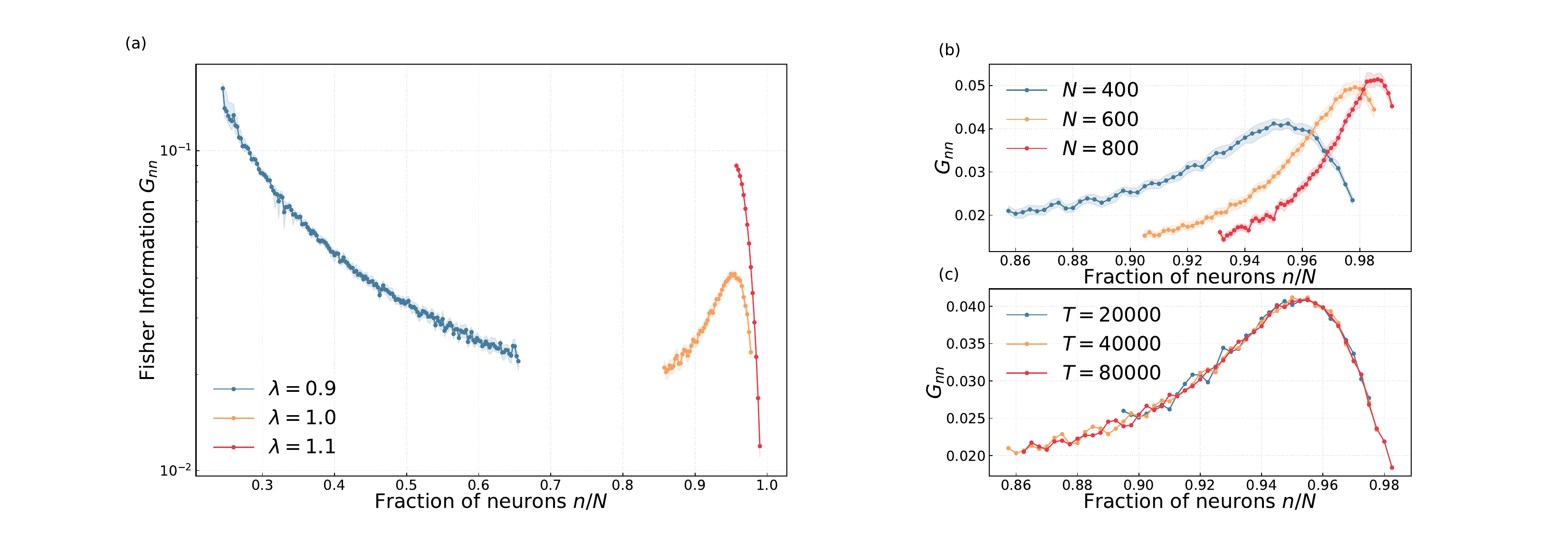}
    \caption{Fisher information of the fraction of unique neurons $G_{nn}$ 
in the spiking neural network. (a) $G_{nn}$ as a function of $n/N$ for 
subcritical ($\lambda=0.9$), critical ($\lambda=1.0$), and supercritical 
($\lambda=1.1$) networks. $N=400$, $T=40000$. Only the critical network 
exhibits a pronounced peak at intermediate $n^*/N$, while subcritical 
and supercritical networks show monotonic decreasing behavior. (b) Spatial 
finite-size effects: increasing $N$ shifts the peak toward larger $n/N$ 
and increases the peak height. $\lambda=1.0$, $T=40000$. (c) Temporal 
finite-size effects: the peak position and height are insensitive to $T$. $\lambda=1.0$, $N=400$. $\tau_{\lambda=0.9}=4, \tau_{\lambda=1.0}=8, \tau_{\lambda=1.1}=10.$}
    \label{fig:fim_of_n}
\end{figure*}

Beyond FIM of the observed branching ratio, here we consider another phenomenological FIM $G_{nn}$, where $n$ is an empirical parameter derived from the microscopic structure of network activity: the number of unique neurons that fire at least once within a time window of length $\tau$. This quantity captures the spatial extent of activity propagation across the network and is directly accessible in experimental recordings. We compute $G_{nn}$ with variable being the spatial-temporal average firing rate $S$ within the time window $S = \frac{1}{\tau}\sum_t S(t),~S(t) = \frac{1}{N}\sum_n s_n(t)$.

As shown in Fig.~\ref{fig:fim_of_n}(a), $G_{nn}$ exhibits qualitatively different behavior across the three critical regimes. For the critical network, $G_{nn}$ displays a pronounced peak at an intermediate value $n^*$, where activity neither saturates the network nor remains confined to a small subpopulation, and small variations in spatial recruitment can trigger large-scale reorganizations of network activity. In the subcritical regime, activity is sparse so that increasing $n$ primarily recruits previously silent neurons, causing changes in $P(S|n)$ that diminish as $n$ grows, yielding a monotonically decreasing $G_{nn}$. For supercritical network, by the similar reasoning activity is near-saturated that further recruitment produces only marginal changes in the distribution, again resulting in a monotonic decrease in $G_{nn}$.

We further analyze the spatial and temporal finite size effects of $G_{nn}$ at criticality $\lambda=1$. As shown in Fig.~\ref{fig:fim_of_n}(b), 
increasing the network size $N$ shifts the peak position toward larger 
$n/N$ and increases the peak height, consistent with the expected 
divergence of susceptibility in the infinite system size limit. In contrast, Fig.~\ref{fig:fim_of_n}(c) shows that the peak position and height of $G_{nn}$ are insensitive to the temporal system size $T$. This is because $n$ and $S$ are both computed within independent fixed-length windows of size $\tau$, so that increasing $T$ only increases the number of time windows without altering the intrinsic network dynamics.

We note that the fundamental difference between $G_{nn}$ and $G_{\sigma\sigma}$ lies in the choice of parameter space: $\sigma=1$ identifies the optimal temporal propagation at criticality, while $n=n^*$ captures its spatial counterpart, the optimal fraction of active neurons that maximizes the sensitivity in critical network. Although FIM is not invariant under reparameterization, both parameterizations reflect maximal sensitivity and large-scale fluctuations, which are fundamental properties of critical systems.

\section{Whole-Brain Model}
\begin{figure*}[!htbp]
    \includegraphics[width=\linewidth]{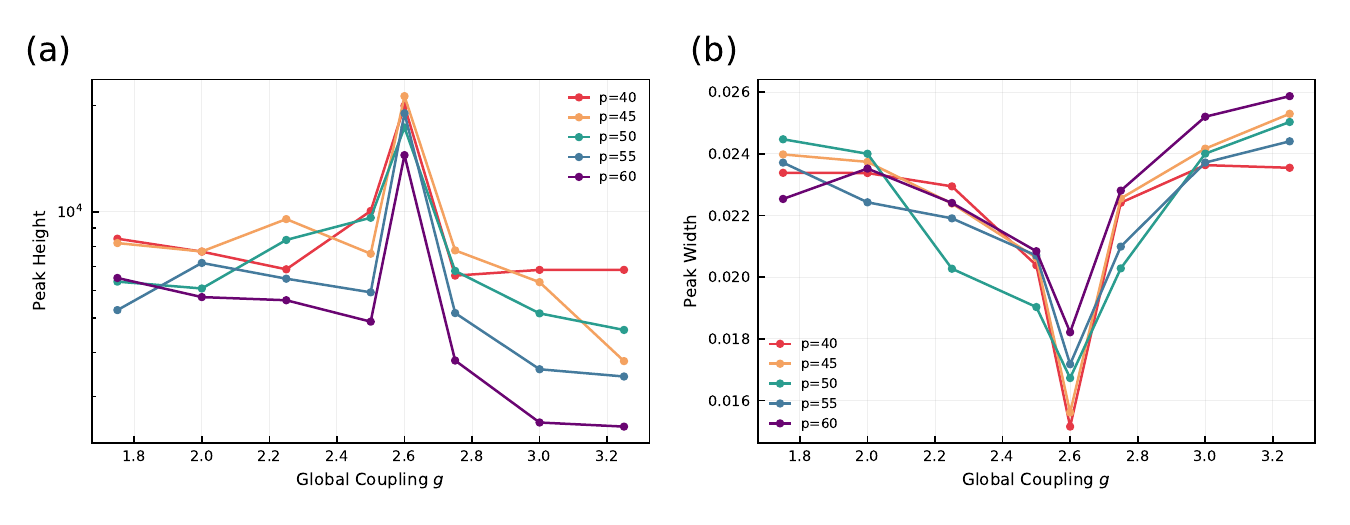}
    \caption{Dependence of the peak height (a) and peak width (b) of the Fisher information $G_{\sigma\sigma}$ on the binarization threshold (the percentile of each region's firing rate distribution used to define active/inactive states) across different coupling strengths $g$. Both metrics are robust to the choice of threshold within a wide range (40th–60th percentile).}
    \label{fig:robustness_of_thres_c90}
\end{figure*}

We assess the robustness of FIM to the binarization threshold, the percentile of each region's firing rate distribution used to define active/inactive states (Fig.~\ref{fig:robustness_of_thres_c90}). Across a wide range of thresholds, the peak height remains highest and the peak width narrowest near the critical point $g_c$, which confirms the FIM $G_{\sigma \sigma}$ is robust to this methodological choice.

\begin{figure*}[!htbp]
    \includegraphics[width=\linewidth]{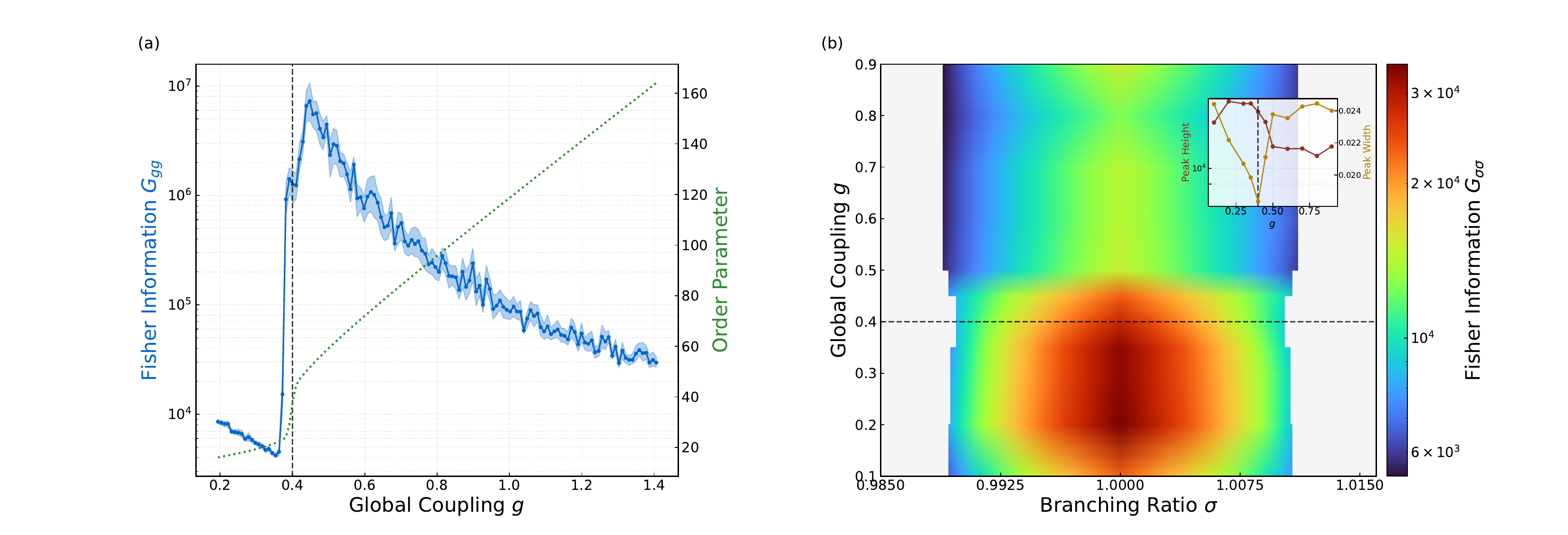}
    \caption{Whole-brain model. (a) Fisher information of the global coupling, $G_{gg}$, for Schaefer100 parcellation with 100 brain regions, showing a pronounced peak near $g_c \approx0.4$.  Dotted line indicates the order parameter $S_v$. (b) Fisher information of the branching ratio, $G_{\sigma\sigma}$, shown as a heat map in the $(g,\sigma)$ plane. The Fisher information peaks at $\sigma=1$ throughout the critical regime, with the peak becoming sharpest and narrowest near $g_c\approx0.4$, as shown in the inset. The light gray region marks values outside the sampled range; only a finite set of $\sigma$ values close to $\sigma=1$ is evaluated.}
    \label{fig:fim_whole_brain_model_C100}
\end{figure*}

\begin{figure*}[!htbp]
    \includegraphics[width=\linewidth]{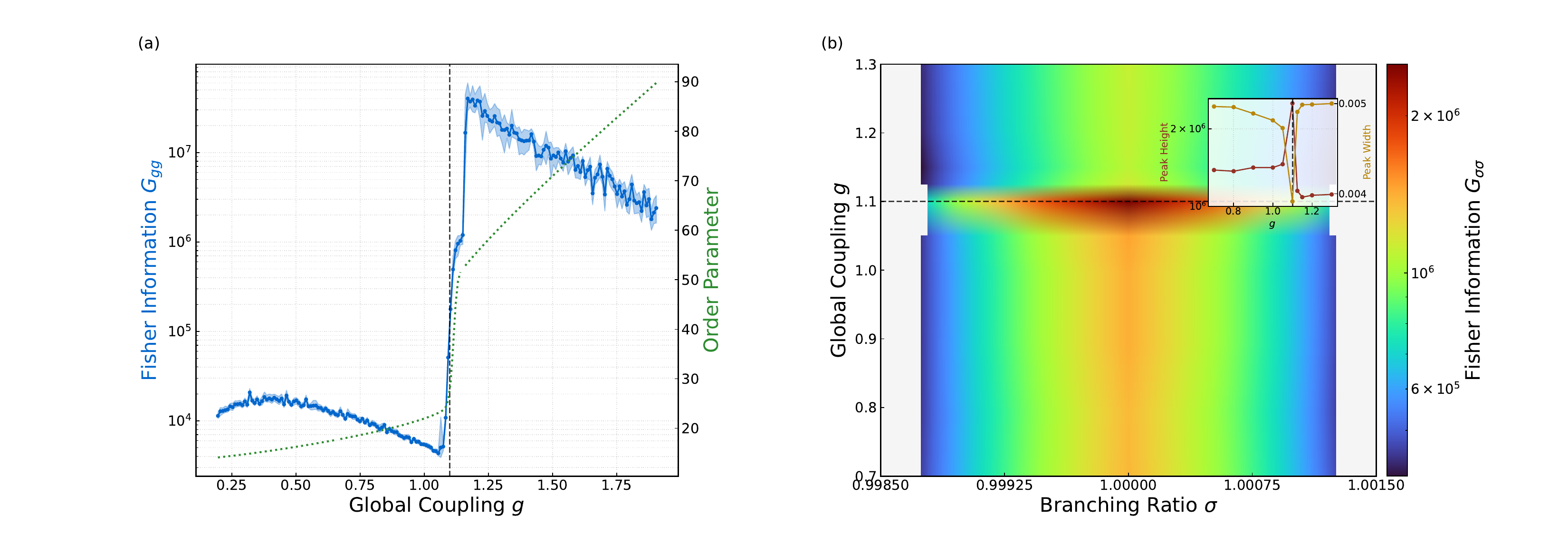}
    \caption{Whole-brain model. (a) Fisher information of the global coupling, $G_{gg}$, for Schaefer1000 parcellation with 1000 brain regions, showing a pronounced peak near $g_c \approx1.1$.  Dotted line indicates the order parameter $S_v$. (b) Fisher information of the branching ratio, $G_{\sigma\sigma}$, shown as a heat map in the $(g,\sigma)$ plane. The Fisher information peaks at $\sigma=1$ throughout the critical regime, with the peak becoming sharpest and narrowest near $g_c\approx1.1$, as shown in the inset. The light gray region marks values outside the sampled range; only a finite set of $\sigma$ values close to $\sigma=1$ is evaluated.}
    \label{fig:fim_whole_brain_model_C1000}
\end{figure*}

Fig.~\ref{fig:fim_whole_brain_model_C100} and Fig.~\ref{fig:fim_whole_brain_model_C1000} applies the two-fold FIM framework to other empirical parcellations with larger brain regions. Again, both FIM of the theoretical control parameter and FIM of observed branching ratio identify the critical region with good accuracy. 

\begin{figure*}[!htbp]
    \includegraphics[width=\linewidth]{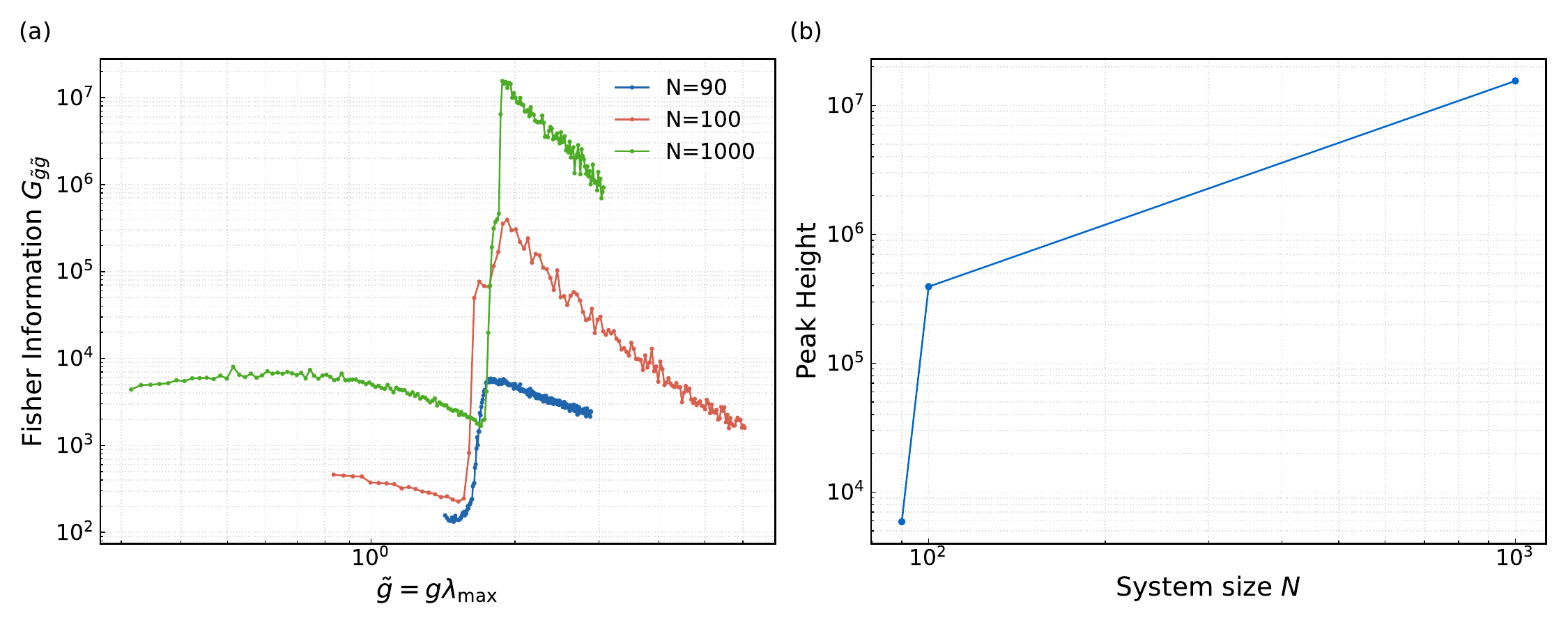}
    \caption{(a) Rescaled Fisher information $G_{\tilde{g} \tilde{g}}$ plotted against the rescaled coupling $g\lambda_{\max}(C)$,
    where $\lambda_{\max}(C)$ is the largest eigenvalue of the structural connectivity. The three peaks collapse onto a common location, indicating that
    $g_c \cdot \lambda_{\max}(C) \approx \text{constant}$ across system
    sizes. (b) Peak height of the rescaled Fisher information as a function of system size $N$, showing that the peak height grows with $N$, consistent with a finite-size signature of criticality.}
    \label{fig:Rescaled_fim_SM}
\end{figure*}

Different structural connectomes yield different critical couplings $g_c$, which complicates direct comparison of Fisher information across system sizes. To enable such a comparison, we
rescale the global coupling using the largest eigenvalue of each connectome,
$\tilde{g} = g \cdot \lambda_{\max}(C)$, and correspondingly reparametrize the Fisher information. Empirically, we find that the rescaled
critical points $g_c \cdot \lambda_{\max}(C)$ collapse onto an approximately common value across
the three connectomes (Fig.~\ref{fig:Rescaled_fim_SM}), allowing us to compare the finite-size behavior of the Fisher information peak across system sizes on a common footing. The reparametrized FIM $G_{\tilde{g} \tilde{g}}$ peak height grows with system size, providing
evidence that this is a genuine finite-size signature of criticality rather than an artifact of a particular connectome's normalization. Note that we do not expect this constant to equal exactly $1$: unlike linear branching-type models, where criticality is defined by $\lambda_{\max}=1$, since the whole-brain model's transfer function and feedback inhibition control~(FIC) introduce additional local factors that shift this constant away
from $1$ in a model-dependent way.

\section{Real Neuronal Data}
\begin{figure}[!htbp]
    \includegraphics[width=\linewidth]{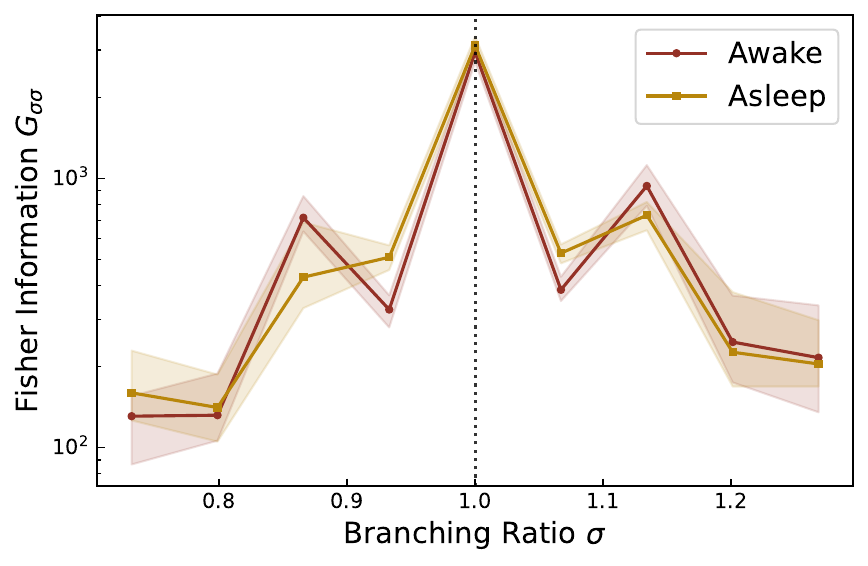}
    \caption{Real neuronal data. Fisher information of the branching ratio $G_{\sigma\sigma}$ computed from real neuronal spike recordings in awake and non-REM sleep states. Peaks at $\sigma = 1$ are observed in both brain states, consistent with criticality. The peak height and width are not significantly distinguishable between states.}
    \label{fig:fim_real_data}
\end{figure}

Having investigated FIM in three systems of increasing complexity, we 
next apply the same methodology to experimental spiking data. We analyze publicly 
available spiking recordings from mice during awake and non-REM sleep 
states, comprising datasets from the Buzsaki Lab with neural ensembles 
ranging from 70 to 119 neurons per session~\cite{petersen_2020_4307883}. 
For each dataset, recordings corresponding to awake and sleep states were 
pooled separately, excluding segments shorter than 60 time steps to 
ensure statistical reliability. We compute $G_{\sigma\sigma}$ separately for awake and sleep states. As shown in Fig.~\ref{fig:fim_real_data}, peaks at $\sigma = 1$ are observed in both states, but the peak height and width remain difficult to distinguish between the two states. This may reflect finite-size limitations in both the number of recorded neurons $N$ and the recording duration $T$. On the other hand, this may reflect the absence of structure-informed weighting: in the whole-brain model, eigenvector weighting derived from the DTI connectome enhances sensitivity to the phase transition by emphasizing hub regions that dominate collective dynamics. For single-neuron spike recordings, however, structural connectivity at the neuronal level is still inaccessible, precluding such weighting and leaving a uniform average as the only option. This suggests that incorporating network structure into the FIM analysis is a key direction for improving the detection of brain state differences in real neuronal data.

\bibliographystyle{unsrt}
\bibliography{mybib}